\documentclass[doublecol]{epl2} 
\usepackage{lineno,hyperref}
\usepackage{graphicx}
\usepackage{amsmath}
\usepackage{amsfonts}
\usepackage{amssymb}
\usepackage{epstopdf}
\usepackage{color}
\usepackage{bm}
\newcommand{\ket}[1]{|#1\rangle}
\newcommand{\bra}[1]{\langle #1|}

\title{Realization of universal nonadiabatic geometric control on decoherence-free qubits in the XY model}

\author{Vahid Azimi Mousolou\inst{1,2}}

\institute{                    
  \inst{1} Department of Mathematics, Faculty of Science, University of Isfahan, Box 81745-163 Isfahan, Iran\\
  \inst{2} School of Mathematics, Institute for Research in Fundamental Sciences (IPM), P. O. Box 19395-5746, Tehran, Iran
}

\abstract{A fundamental requirement of quantum information processing is the protection from the adverse effects of decoherence and noise. Decoherence-free subspaces and geometric processing are important steps of quantum information protection. Here, we provide a new experimentally feasible scheme to combine decoherence-free subspaces with nonadiabatic geometric manipulations to attain a universal quantum computation. The proposed scheme is different from previous proposals and is based on the typical XY interaction coupling, which can be set up in various nano-engineered systems and therefore open up for realization of nonadiabatic holonomic quantum computation in decoherence-free subspaces.}

\begin{document}

\maketitle

\section{Introduction}

Quantum technology suffers from the fragile nature of quantum states. In the field of quantum information processing different strategies and methods have been employed to overcome this issue. For instance, quantum error-correction methods \cite{niellsen10}, decoherence free subspaces \cite{zanardi1997a, lidar1998, lidar2003}, noise-less subsystems \cite{zanardi1997, knill2000}, dynamical decoupling \cite{viola1998, viola1999}, topological and geometric approaches \cite{kitaev2003, zanardi99, sjoqvist12} to store, maintain and process the quantum information are among the key tools. Since each of these methods plays an exclusively crucial role in quantum computation and information science, combination of numbers of these methods to hopefully achieve a practical fault resistant hybrid mechanism for quantum information processing has always been at the center of research attentions \cite{zanardi2003, wu2005, zhang2006, cen2006, feng2009, oreshkov2009, Xu2012, Xu2014a, Liang2014, Xue2015, Zhou2015, Xue2016, zhao2017, Zhang2014, Xu2014b}. 

A decoherence-free subspace (DFS) is a symmetric subspace of a system's Hilbert space, which is invariant by collective decohering processes generated in certain system's interactions with the dephasing environment. The idea of DFSs started with observations in a study of pure dephasing on two qubits that have identical interactions with the environment \cite{palma1996}. Subsequently, this phenomenon was developed further using different methods \cite{zanardi1997a, lidar1998, lidar2003, duan1997, duan1998, zanardi1997, zanardi1998, lidar1999} and found its application in fault-tolerant quantum computation \cite{zanardi1998a, zanardi1999a, kempe2001}. The explicit theoretical demonstrations followed by experimental verifications \cite{kwiat2000, kielpinski2001, viola2001, fortunato2002, pushin2011} revealed the potential of DFSs in protecting fragile quantum information against decoherences and identified them as a major component in the construction of a scalable quantum computer \cite{kielpinski2002, brown2003, aolita2007, monz2009}. 

Geometric or holonomic quantum computation is the idea of building up quantum processors based on quantum geometric phases/holonomies. This model initially proposed based on quantum holonomy accompanying adiabatic evolutions \cite{zanardi99, ekert2000, duan2001, xiang-bin2001, faoro2003, solinas2003} and then involved nonadiabatic evolutions \cite{sjoqvist12, zhu2002, zhu2003a, zhu2003b, sjoqvist2012, mousolou2014, mousolou2017a}. Inherent robustness of quantum geometric phases \cite{Pachos2001, Solinas2012} due to its conceptually natural relation to the geometric description of quantum systems \cite{bohm2003, dariusz2004}, has turned holonomic quantum computation into one of the key approaches to achieve fault-tolerant quantum computation. Holonomic gates have been implemented in various experimental settings, such as NMR \cite{jones2000, du2006, feng2013}, ion traps \cite{leibfried2002, toyoda2013}, superconducting transmon \cite{Abdumalikov2013}, NV-centers in diamond \cite{Arroyo-Camejo2014, Zu2014}, and other solid-state systems \cite{tian2009}. 

To take the potential advantages of both decoherence-free subspaces and holonomic quantum computation methods in building up fault-tolerant architectures for quantum computing, several hybrid proposals have been put forth \cite{wu2005, zhang2006, cen2006, feng2009, Xu2012, Xu2014a, Liang2014, Xue2015, Zhou2015, Xue2016, zhao2017} for adiabatic as well as nonadiabatic regimes. However, the need for control of complex qubit interactions in the proposed schemes has made it a challenging task to fully demonstrate these architectures in lab. Here, we present a new practically feasible scheme for implementation of nonadiabatic holonomic quantum computation in decoherence-free subspaces. This scheme is only based on typical one-dimensional $XY$ interaction model, which is an important model in quantum information science and has been realized in diverse experimental settings \cite{niederberger2010, alvarez2010, rao2014, imamoglu1999, barredo2015, ping2013, zhu2011, struck2011, kosior2013, berloff2017, tamate2016, takeda2017}. We use two physical qubits to encode a logical qubit into a two dimensional decoherence-free subspace. Then, by embedding the logical qubit into a three dimensional decoherence-free subspace and constructing a lambda structure through an anisotropic three bodies $XY$ interaction Hamiltonian, we implement a universal set of nonadiabatic holonomic single-qubit gates on our logical qubit. Next, we create a double lambda structure in an anisotropic XY interaction coupling between two logical qubits and demonstrate a family of nonadiabatic two-qubit entangling holonomic operations, which includes entangling gate equivalent to CNOT gate.

The paper is organized as follows: We begin with a brief introduction of nonadiabatic holonomic gates in Sec.~\ref{Nonadiabatic holonomic gates}. In Sec.~\ref{Model system}, we encode the logical qubit into a decoherence-free subspace and establish a universal set of nonadiabatic holonomic single-qubit gates. We continue by constructing a family of nonadiabatic two-qubit holonomic gates and evaluating their entangling nature in Sec.~\ref{Two-qubit entangling gate}. The paper is summarized in Sec.~\ref{summary}. 

\section{Nonadiabatic holonomic gates}
\label{Nonadiabatic holonomic gates}
Below, we shall briefly address nonadiabatic holonomic gates \cite{sjoqvist12}, which are based on non-adiabatic non-Abelian geometric phase \cite{anandan88}.

Consider a quantum system described by a (time dependent) Hamiltonian $H$ operating on the corresponding $n$-dimensional Hilbert state space $\mathcal{H}$. Suppose $W$ is a $k$-dimensional subspace of $\mathcal{H}$, which undergoes the Schr\"odinger time evolution
\begin{eqnarray}
\mathcal{C}\ :\ [0,\tau]\ni t\rightarrow W(t)
\label{path-Grass}
\end{eqnarray}
in a cyclic manner, i.e., $W(\tau)=W(0)=W$ for some time $\tau$. Despite the fact that the evolution $\mathcal{C}$ represents a cyclic physical process,  
the evolution of a single pure quantum state $\ket{\psi(t)}$ about $\mathcal{C}$ need not to be cyclic, or, more precisely, the initial, $\ket{\psi(0)}$, and final ,$\ket{\psi(\tau)}$, states in the base subspace $W$ do not necessarily coincide. However, there always exists a unitary operator $U(\tau)$ such that $\ket{\psi(\tau)}=U(\tau)\ket{\psi(0)}$. The unitary $U(\tau)$ is just the time evolution operator $\mathcal{U}(0,\tau)=e^{-\frac{i}{\hbar}\int_{0}^{\tau}Hdt}$ projected on the base subspace $W$. This unitary transformation can be evaluated as  
\begin{eqnarray}
U(\tau)=\textbf{P}e^{i\int_{0}^{\tau} (A-D)dt}, 
\label{eq:gate}
\end{eqnarray}
where $\textbf{P}$ is the time ordering operator and 
\begin{eqnarray}
A_{ab}=i\bra{\tilde{\psi}_a(t)}\frac{d}{dt}\ket{\tilde{\psi}_b(t)}\ \ \&\ \ D_{ab}=\frac{1}{\hbar}\bra{\tilde{\psi}_a(t)}H\ket{\tilde{\psi}_b(t)}\nonumber\\
\end{eqnarray}
for a given once differentiable set of $k$ orthonormal basis state vectors $\tilde{B}(t)=\{\ket{\tilde{\psi}_a(t)},\ a=1,...,k\}$ spanning the subspace $W(t)$ at each time such that
$\ket{\tilde{\psi}_a(\tau)}=\ket{\tilde{\psi}_a(0)}$ for each $a$. The $\tilde{B}(t)$ is sometimes referred to as a one parameter family of $k$-frames. 

Note that, the unitary transformation $U(\tau)$ in Eq. (\ref{eq:gate}) is composed of two parts. One is given by the potential $A$, which transforms as a proper gauge potential under a gauge transformation and thus gives rise to a gauge-covariant geometric contribution to the operator $U(\tau)$ \cite{anandan88}. While, the other is given by the matrix $D$, which is merely given by the dynamics of the system described by the Hamiltonian $H$. The contribution by the gauge potential $A$, here denoted as
\begin{eqnarray}
U(\mathcal{C})=\textbf{P}e^{i\int_{0}^{\tau} Adt}, 
\end{eqnarray}
is known as nonadiabatic quantum holonomy of the evolution $\mathcal{C}$. 

The geometric unitary operator, $U(\mathcal{C})$, would realize a nonadiabatic holonomic gate provided that
\begin{itemize}
\item[i.] The base subspace $W$ represents a system of qubits.
\item[ii.] The subspace $W$ is evolved cyclically about a loop $\mathcal{C}$ in the whole state space $\mathcal{H}$.
\item[iii.]  Along the evolution $\mathcal{C}$, the dynamical potential $D$ vanishes. This is equivalent to that the projected Hamiltonian at each time on the subspace $W(t)$
along the evolution $\mathcal{C}$ vanishes.
\end{itemize}

Geometrically speaking, the loop $\mathcal{C}$ in fact resides in the Grassmannian manifold $\mathcal{G}(n,k)$, the space of all $k$-dimensional subspaces of the $n$-dimensional Hilbert space $\mathcal{H}$. Moreover, the nonadiabatic quantum holonomy $U(\mathcal{C})$ depends only on the loop $\mathcal{C}$ and geometric structure of the $U(k)$-principal bundle $\Gamma=(\mathcal{S}(n,k), \mathcal{G}(n,k)), \pi, U(k))$. Here, the Stiefel manifold $\mathcal{S}(n,k)$ is the space of all $k$-frames in $\mathcal{H}$, $\pi$ is the natural projection operator that maps each $k$-frame to the corresponding spanned subspace, and $U(k)$ is the unitary group of degree $k$  \cite{anandan88}.

\section{Single qubit gate}
\label{Model system}

\begin{figure}[h]
\centering
\includegraphics[width=70mm,height=25mm]{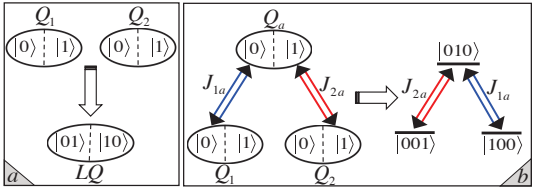}
\caption{(Color online) Panel ($a$): Schematic diagram of the logical qubit (LQ) encoded into the two dimensional $\text{DFS}_{2}$ of two physical qubits, $Q_{1}$ and $Q_{2}$, as $\ket{0_{L}}=\ket{01}$ and $\ket{1_{L}}=\ket{10}$. Panel ($b$): The two physical qubits are coupled through an intermediate ancilla qubit labeled as $Q_{a}$ in order to incorporate the logical qubit into the three dimensional $\text{DFS}_{3}$ spanned by the encoded basis states $\ket{0_{L}}\equiv\ket{001}$, $\ket{a}\equiv\ket{010}$ and $\ket{1_{L}}\equiv\ket{100}$. The ancilla qubit allows us to have a desirable lambda dynamic for the logical qubit.}
\label{fig:DFSQ}
\end{figure}

As depicted in Fig. \ref{fig:DFSQ} (a), we consider a logical qubit (LQ) representted by the two dimensional collective dephasing DFS  \cite{lidar1998, lidar2003}
\begin{eqnarray}
\text{LQ}\equiv\text{DFS}_{2}=\text{Span}\{\ket{01},\ket{10}\}
\label{eq: DFS2}
\end{eqnarray}
of  two physical qubits, $Q_{1}$ and $Q_{2}$. Let us define encoded computational basis states for our logical qubit as
\begin{eqnarray}
\ket{0_{L}}=\ket{01}\ \ \ \ \ \ \&\ \ \ \ \ \ \ket{1_{L}}=\ket{10}.
\label{eq: encodedBS}
\end{eqnarray}

To introduce a qubit dynamic, we embed our logical qubit into a higher dimensional DFS, namely 
\begin{eqnarray}
\text{DFS}_{3}=\text{Span}\{\ket{001},\ket{010}, \ket{100}\},
\label{eq: DFS3}
\end{eqnarray} 
by, as shown in Fig. \ref{fig:DFSQ}(b), coupling the two physical qubits, $Q_{1}$ and $Q_{2}$, through an intermediate auxiliary physical qubit,  $Q_{a}$, via 
the anisotropic XY interaction Hamiltonian in local magnetic fields
\begin{eqnarray}
H_{1}&=& \frac{J_{1a}}{2}[\sigma_{x}^{(1)}\sigma_{x}^{(a)}+\sigma_{y}^{(1)}\sigma_{y}^{(a)}]+\frac{J_{2a}}{2}[\sigma_{x}^{(a)}\sigma_{x}^{(2)}+\sigma_{y}^{(a)}\sigma_{y}^{(2)}]\nonumber\\
&&+ B(\sigma_{z}^{(1)}+\sigma_{z}^{(2)}),
\label{eq:H}
\end{eqnarray}
where $J_{1a}$, $J_{2a}$ denote the exchange coupling strengths between adjacent qubits, $B$ denotes the strength of local magnetic fields, and $\sigma_{x}^{(j)}$, $\sigma_{y}^{(j)}$, $\sigma_{z}^{(j)}$ are the standard Pauli operators of the $j$th qubit. 
 
Since the Hamiltonian in Eq. (\ref{eq:H}) commutes with total $z$-component operator 
\begin{eqnarray}
\sigma_{x}^{(\text{tot})}=\sigma_{z}^{(1)}+\sigma_{z}^{(a)}+\sigma_{z}^{(2)},
\end{eqnarray}
the three dimensional DFS in Eq. (\ref{eq: DFS3}) is an invariant subspace by $H_{1}$. 
Thus, the dynamic of our logical qubit, which is now equivalently represented by the 
subspace $\text{Span}\{\ket{0_{L}}\equiv\ket{001}, \ket{1_{L}}\equiv\ket{100}\}$ 
in $\text{DFS}_{3}$ occurs only in $\text{DFS}_{3}$. The Hamiltonian in Eq. (\ref{eq:H}) in the effective subspace $\text{DFS}_{3}$ takes the following effective lambda form 
\begin{eqnarray}
H_{\text{eff}}=\left(
\begin{array}{ccc}
 0 &   0 & J_{2a}   \\
0  &  0 & J_{1a}   \\
J_{2a}  & J_{1a}  & 2B
\end{array}
\right),
\label{eq:effH}
\end{eqnarray}
with respect to the ordered basis $\{\ket{0_{L}}, \ket{1_{L}}, \ket{a}\equiv\ket{010}\}$ (see Fig. \ref{fig:DFSQ}(b)).

Using the following parameterization 
\begin{eqnarray}
(J_{1a}, J_{2a}, B)=\omega(\sin\phi\cos\frac{\theta}{2},\  \sin\phi\sin\frac{\theta}{2},\  \cos\phi)
\end{eqnarray}
with $\omega=\sqrt{J_{1a}^{2}+J_{2a}^{2}+ B^{2}}$, the effective energy eigenstates are obtained as 
\begin{eqnarray}
\ket{d}&=&\cos\frac{\theta}{2}\ket{0_{L}}-\sin\frac{\theta}{2}\ket{1_{L}}\nonumber\\
\ket{b_{1}}&=&\cos\frac{\phi}{2}\ket{a}+\sin\frac{\phi}{2}\ket{b}\nonumber\\
\ket{b_{2}}&=&\sin\frac{\phi}{2}\ket{a}-\cos\frac{\phi}{2}\ket{b},
\end{eqnarray}
where $\ket{b}=\sin\frac{\theta}{2}\ket{0_{L}}+\cos\frac{\theta}{2}\ket{1_{L}}$, corresponding to energies $E_{d}=0,\ E_{b_{1}}=\omega(\cos\phi+1)$, and $\ E_{b_{2}}=\omega(\cos\phi-1)$. Consequently, we evaluate the time evolution operator, $\mathcal{U}(0, \tau)=e^{-\frac{i}{\hbar}\int_{0}^{\tau}H_{\text{eff}}dt}$, as
\begin{eqnarray}
\mathcal{U}(0,\tau)&=&\ket{d}\bra{d}+\nonumber\\
&&e^{-i\frac{\omega\tau}{\hbar}\cos\phi}\{[\cos\frac{\omega\tau}{\hbar}-i\sin\frac{\omega\tau}{\hbar}\cos\phi]\ket{a}\bra{a}\nonumber\\
&&\ \ \ \ \ \ \ \ \ \ \ \ \ \ [\cos\frac{\omega\tau}{\hbar}+i\sin\frac{\omega\tau}{\hbar}\cos\phi]\ket{b}\bra{b}\nonumber\\
&&\ \ \ \ \ \ \ \ \ \ \ \ \ \ -i\sin\phi\sin\frac{\omega\tau}{\hbar}[\ket{a}\bra{b}+\ket{b}\bra{a}]\}.\nonumber\\
\label{eq:t-evolution}
\end{eqnarray}

In order to achieve single qubit gates on our logical qubit, $LQ\equiv\text{Span}\{\ket{0_{L}}, \ket{1_{L}}\}$, we need to evolve the qubit subspace cyclically in $\text{DFS}_{3}$, i.e., we must let the system evolve within an appropriate time interval, $[0,\tau]$, so that at the final time, $\tau$, the qubit subspace is mapped back into itself by the time evolution operator $\mathcal{U}(0,\tau)$. Since 
\begin{eqnarray}
\text{Span}\{\ket{0_{L}}, \ket{1_{L}}\}=\text{Span}\{\ket{d}, \ket{b}\},
\end{eqnarray}
the last term in Eq (\ref{eq:t-evolution}) implies that we would have a cyclic evolution for the logical qubit in $\text{DFS}_{3}$ if we assume a time $\tau$ with respect to the exchange couplings and local magnetic fields strengths such that 
\begin{eqnarray}
\frac{\omega\tau}{\hbar}=m\pi,
\end{eqnarray}
for an integer $m$. In this case, we obtain single qubit gate
\begin{eqnarray}
U(\mathcal{C})=e^{-i\frac{\gamma}{2}}\{\cos\frac{\gamma}{2}\hat{1}-i\sin\frac{\gamma}{2}[\sin\theta X-\cos\theta Z]\}
\label{SQG}
\end{eqnarray}
where $\gamma=m\pi(\cos\phi+1)$. The $\hat{1}, X$, and  $Z$ are, respectively, identity, Pauli $X$ and Pauli $Z$ single qubit gates. The $U(\mathcal{C})$ follows from projecting the time evolution operator $\mathcal{U}(0,\tau)$ on the logical qubit subspace. Here, $\mathcal{C}$ denotes the cyclic path in $\mathcal{G}(3,2)$ about which the logical qubit subspace $\text{LQ}\equiv\text{DFS}_{2}$ evolves.   

\begin{figure}[t]
\centering
\includegraphics[width=60mm,height=40mm]{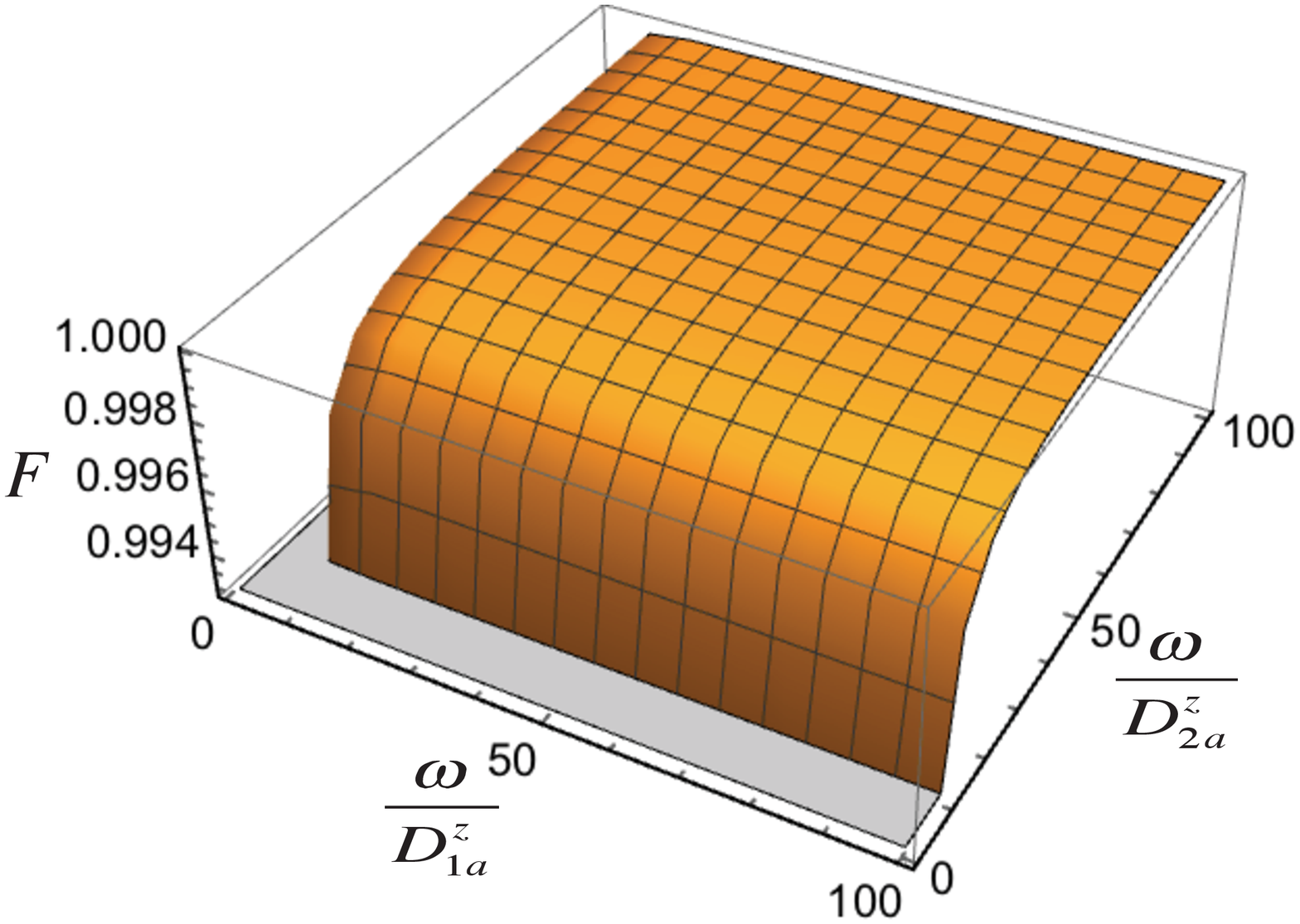}
\includegraphics[width=60mm,height=40mm]{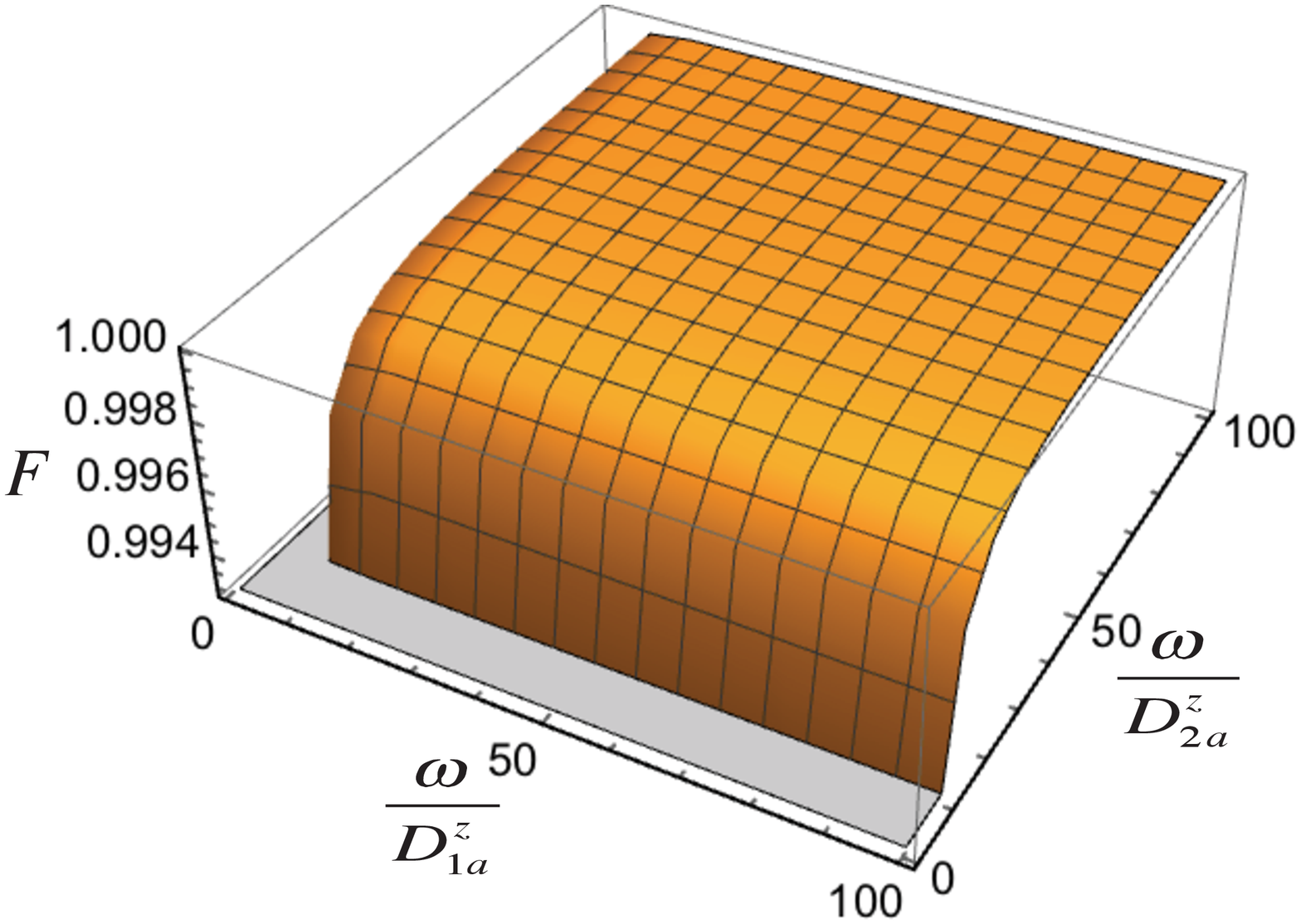}
\caption{(Color online) Fidelity, $F$, of the nonadiabatic holonomic Hadamard (upper panel) and $\frac{\pi}{8}$ (lower panel) gates in decoherence-free subspace against dimensionless parameters $\frac{\omega}{D^{z}_{1a}}$ and $\frac{\omega}{D^{z}_{2a}}$ corresponding to the Dzyalozhinsky-Moriya spin-orbit interaction contribution to the system. We have used the antisymmetric Dzyalozhinsky-Moriya term 
$\frac{D^{z}_{1a}}{2}[\sigma_{x}^{(1)}\sigma_{y}^{(a)}-\sigma_{y}^{(1)}\sigma_{x}^{(a)}]+\frac{D^{z}_{2a}}{2}[\sigma_{x}^{(a)}\sigma_{y}^{(2)}-\sigma_{y}^{(a)}\sigma_{x}^{(2)}]$.}
\label{fig:Fid-SQG}
\end{figure}

Let us pursue with some remarks on the gate operator, $U(\mathcal{C})$, given in Eq. (\ref{SQG}):
\begin{itemize}
\item The $U(\mathcal{C})$ is nonadiabatic holonomic gate since it meets all the criterias listed at the end of Sec. \ref{Nonadiabatic holonomic gates}. Firstly, it corresponds to cyclic evolution $\mathcal{C}$ of the logical qubit subspace $\text{LQ}\equiv\text{DFS}_{2}$ in the whole effective state space $\text{DFS}_{3}$, and secondly, 
\begin{eqnarray}
\bra{x}e^{\frac{i}{\hbar}\int_{0}^{t}H_{\text{eff}}ds}H_{\text{eff}}e^{-\frac{i}{\hbar}\int_{0}^{t}H_{\text{eff}}ds}\ket{y}=\bra{x}H_{\text{eff}}\ket{y}=0\nonumber\\
\end{eqnarray}
at each time $t$ along this evolution for $x,y=0_{L},1_{L}$. The latter indicates that there is no dynamical contribution in $U(\mathcal{C})$. Thus, the $U(\mathcal{C})$ is actually the nonadiabatic quantum holonomy associated with the evolution of the logical qubit subspace about $\mathcal{C}$.
\item The $U(\mathcal{C})$ provides a universal set of single qubit gates since it represents the rotation by an arbitrary angle $\gamma$ about an arbitrary axis $\hat{n}=(\sin\theta,\ 0,\ -\cos\theta)$ in the $xz$ plane, i.e., $U(\mathcal{C})\equiv\mathcal{R}_{\hat{n}}(\gamma)$  \cite{shim2013}.
\item Compared to the previous proposals \cite{Xu2012, zhao2017}, the gate $U(\mathcal{C})$ gets its universality and geometric nature without taking into account an external control on the Dxzialoshinsky-Moriya spin orbit interaction term in the Hamiltonian of Eq. (\ref{eq:H}). Instead,  we consider the Dxzialoshinsky-Moriya term, which is an intrinsic property of the system, very small and probably not easy to control externally in practice, as a perturbative noise to the system. Figure \ref{fig:Fid-SQG} shows remarkably high tolerance of the $U(\mathcal{C})$ gate against this noise contribution.
\end{itemize}

\section{Two-qubit entangling gate}
\label{Two-qubit entangling gate}
To illustrate the full computational power of the proposed geometric scheme in decoherence-free subspaces, here we discuss the realization of nonlocal two-qubit holonomic gates. 

\begin{figure}[h]
\centering
\includegraphics[width=70mm,height=25mm]{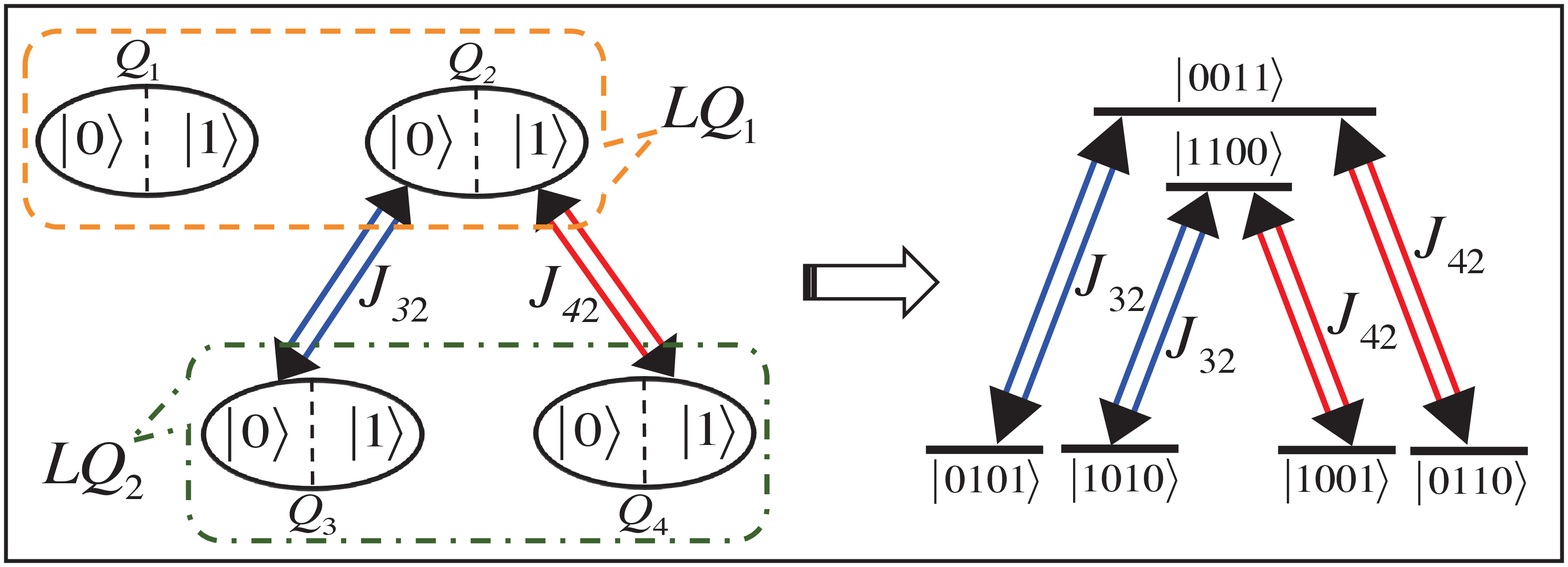}
\caption{(Color online) Schematic diagram of two pairs of physical qubits $\{Q_{1}, Q_{2}\}$ and $\{Q_{3}, Q_{4}\}$, respectively, representing two decoherence-free logical qubits labeled as  $LQ_{1}$ and $LQ_{2}$. An XY coupling between three of these four physical qubits, say here $Q_{3}$, $Q_{2}$, $Q_{4}$, leads to a double lambda dynamic for the two logical qubits computational state space in the six dimensional decoherence-free subspace $\text{DFS}_{6}$. This dynamic allows us to implement a geometric two-qubit entangling gate between logical qubits with a desirable entangling power.}
\label{fig:DFS2Q}
\end{figure}

As seen from the former section and also shown in Fig. \ref{fig:DFS2Q}, two logical qubits would be represented with four or precisely two pairs of physical qubits. For the sake of clarity,  we label the physical qubits denoting each logical qubit as follows:
\begin{eqnarray}
LQ_{1}&\leftarrow&\{Q_{1}, Q_{2}\}\nonumber\\
LQ_{2}&\leftarrow&\{Q_{3}, Q_{4}\}.
\end{eqnarray}

We consider the coupling between the two logical qubits introduced by the following XY interaction Hamiltonian

\begin{eqnarray}
H_{2}=\frac{J_{32}}{2}[\sigma_{x}^{(3)}\sigma_{x}^{(2)}+\sigma_{y}^{(3)}\sigma_{y}^{(2)}]+\frac{J_{42}}{2}[\sigma_{x}^{(4)}\sigma_{x}^{(2)}+\sigma_{y}^{(4)}\sigma_{y}^{(2)}],\nonumber\\
\label{eq:H2}
\end{eqnarray}
where $J_{32}$ and $J_{42}$ are the exchange coupling strengths between the corresponding qubits.
Note that, similar to the single qubit case, the Hamiltonian in Eq. (\ref{eq:H2}) commutes with total $z$-component operator 
\begin{eqnarray}
\sigma_{x}^{(\text{tot})}=\sigma_{z}^{(1)}+\sigma_{z}^{(2)}+\sigma_{z}^{(3)}+\sigma_{z}^{(4)},
\end{eqnarray}
which indicates that the six dimensional DFS 
\begin{eqnarray}
\text{DFS}_{6}=\text{Span}\{\ket{0101}, \ket{1010}, \ket{0110}, \ket{1001}, \ket{0011}, \ket{1100}\}\nonumber\\
\end{eqnarray}
including the two logical qubits computational state space encoded as 
\begin{eqnarray}
\ket{0_{L}0_{L}}&=&\ket{0101},\ \ \ \ \ket{1_{L}1_{L}}= \ket{1010}\nonumber\\
\ket{0_{L}1_{L}}&=&\ket{0110},\ \ \ \ \ket{1_{L}0_{L}}=\ket{1001}
\end{eqnarray}
remains invariant under the the Hamiltonian in Eq. (\ref{eq:H2}). Thus, the dynamics of the two logical qubits take place only in $\text{DFS}_{6}$. Geometrically speaking, an evolution of the two logical qubits computational state space would correspond to a path in the Grassmannian manifold $\mathcal{G}(6,4)$. Restricting the Hamiltonian in Eq. (\ref{eq:H2}) into the effective subspace $\text{DFS}_{6}$, we obtain the effective Hamiltonian of the form
\begin{eqnarray}
\tilde{H}_{\text{eff}}=\left(
\begin{array}{cccccc}
 0 & 0 & 0 & 0& J_{32} & 0   \\
 0 & 0 & 0 & 0& 0 & J_{32}   \\
 0 & 0 & 0 & 0& J_{42} & 0   \\
 0 & 0 & 0 & 0& 0 & J_{42}   \\
J_{32} & 0 & J_{42} & 0& 0 & 0   \\
 0 & J_{32} & 0 & J_{42}& 0 & 0   
\end{array}
\right)
\end{eqnarray}
in the ordered basis $\{\ket{0101}, \ket{1010}, \ket{0110}, \ket{1001}, \ket{0011}$ $, \ket{1100}\}$. As depicted in Fig. \ref{fig:DFS2Q}, this Hamiltonian benefits actually from a double lambda form 
\begin{eqnarray}
\tilde{H}_{\text{eff}}=\left(
\begin{array}{ccc}
 0 & 0 &  J_{32}    \\
 0 & 0 &  J_{42}    \\
J_{32} & J_{42} & 0   
\end{array}
\right)\otimes
\left(\begin{array}{cc}
 1 & 0  \\
 0 & 1  
\end{array}
\right).
\end{eqnarray}

Note that, the first term on the right hand side has the same form as the Hamiltonian in Eq. (\ref{eq:effH}). Therefore, following the analysis of the former section we observe that the two logical qubits state space evolves cyclically about a loop in the Grassmannian manifold $\mathcal{G}(6,4)$ if we assume a time interval, $[0,\tilde{\tau}]$, for the evolution such that 
\begin{eqnarray}
\frac{\tilde{\omega}\tilde{\tau}}{\hbar}=\tilde{m}\pi
\end{eqnarray}
for an integer $\tilde{m}$. Here, we have $\tilde{\omega}=\sqrt{J_{32}^{2}+J_{42}^{2}}$. In the case of an odd integer $\tilde{m}$, the projection of the time evolution operator, 
$\mathcal{U}(0, \tilde{\tau})=e^{-\frac{i}{\hbar}\int_{0}^{\tilde{\tau}}\tilde{H}_{\text{eff}}dt}$, on the two logical qubits state space leads the nonlocal conditional two-qubit gate 
\begin{eqnarray}
U(\tilde{\mathcal{C}})&=&\left(
\begin{array}{cccc}
 \cos\tilde{\theta} & -\sin\tilde{\theta} & 0 & 0   \\
 -\sin\tilde{\theta} & - \cos\tilde{\theta}& 0 & 0   \\
 0 & 0 & -\cos\tilde{\theta} & -\sin\tilde{\theta}   \\
 0 & 0 & -\sin\tilde{\theta} & \cos\tilde{\theta} 
\end{array}
\right)\nonumber\\
&\equiv&\ket{0_{L}}\bra{0_{L}}\otimes\mathcal{R}_{\hat{\nu}_{0}}(\pi)+\ket{1_{L}}\bra{1_{L}}\otimes\mathcal{R}_{\hat{\nu}_{1}}(\pi)\nonumber\\
\label{eq:2qgate}
\end{eqnarray}
represented in the ordered computations basis $\{\ket{0_{L}0_{L}}, \ket{0_{L}1_{L}}, \ket{1_{L}0_{L}}, \ket{1_{L}1_{L}}\}$. Here, we have used 
\begin{eqnarray}
J_{32}=\tilde{\omega}\sin\frac{\tilde{\theta}}{2}\ \ \ \ \&\ \ \ \ J_{42}=\tilde{\omega}\cos\frac{\tilde{\theta}}{2}.
\end{eqnarray}
In Eq. (\ref{eq:2qgate}), $\mathcal{R}_{\hat{\nu}_{0}}(\pi)$ and $\mathcal{R}_{\hat{\nu}_{1}}(\pi)$ are conditional $\pi$ rotations of the second qubit about the corresponding axes 
\begin{eqnarray}
\hat{\nu}_{0}=(\sin\tilde{\theta}, 0, -\cos\tilde{\theta})\ \ \ \&\ \ \ \hat{\nu}_{1}=(\sin\tilde{\theta}, 0, \cos\tilde{\theta}).\ \ \ 
\end{eqnarray}

Two important features of the two-qubit gate, $U(\tilde{\mathcal{C}})$, in Eq. (\ref{eq:2qgate}), which we would like to stress here, are that it is a holonomic and entangling gate.
The holonomic nature follows from the fact that the $U(\tilde{\mathcal{C}})$ corresponds to a cyclic path, $\tilde{\mathcal{C}}$, in the Grassmannian manifold $\mathcal{G}(6,4)$, about which the two logical qubits state space evolves and along this evolution there is no dynamical contributions since 
\begin{eqnarray}
\bra{xy}e^{\frac{i}{\hbar}\int_{0}^{t}\tilde{H}_{\text{eff}}ds}\tilde{H}_{\text{eff}}e^{-\frac{i}{\hbar}\int_{0}^{t}\tilde{H}_{\text{eff}}ds}\ket{pq}=\bra{xy}\tilde{H}_{\text{eff}}\ket{pq}=0,\nonumber\\
\end{eqnarray}
at each time $t\in[0, \tilde{\tau}]$ for $p, q, x, y=0_{L}, 1_{L}$. According to the sec. \ref{Nonadiabatic holonomic gates}, the $U(\tilde{\mathcal{C}})$ is the nonadiabatic quantum holonomy associated with the cyclic evolution $\tilde{\mathcal{C}}$. 

\begin{figure}[h]
\centering
\includegraphics[width=80mm,height=30mm]{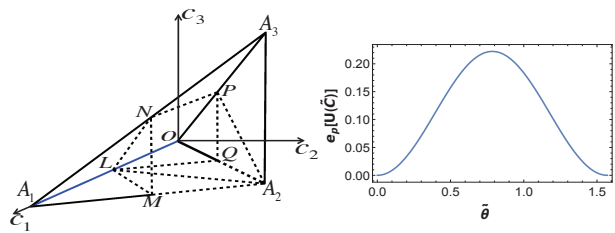}
\caption{(Color online) Left panel: symmetry reduced representation of the geometric two-qubit gate $U(\tilde{\mathcal{C}})$ on the so called tetrahedron Weyl chamber classifying non-local two-qubit gate operations. The $U(\tilde{\mathcal{C}})$ gate covers the whole edge $OA_{1}$, shown in blue, for different values of $0<\tilde{\theta}<\pi/2$. For $\tilde{\theta}=\pi/4$, the gate $U(\tilde{\mathcal{C}}; \tilde{\theta}=\pi/4)$ represents the point $L$ on the Weyl chamber, which corresponds to the CNOT equivalence class. Right panel: Entangling power, $e_{p}[U(\tilde{\mathcal{C}})]$, as a function of the control parameter $\tilde{\theta}$. Any entangling power is achieved by adjusting the exchange couplings $J_{32}$ and $J_{42}$ so that $0<\tilde{\theta}=2\arctan\frac{J_{32}}{J_{42}}<\pi/2$. At $\tilde{\theta}=\pi/4$ the maximum gate entangling power of $2/9$ is obtained.}
\label{fig:EP-WC}
\end{figure}

To clarify the entangling nature, we have evaluated the local invariants \cite{balakrishnan09} for the $U(\tilde{\mathcal{C}})$ as
\begin{eqnarray}
G_{1}=\cos^{2}2\tilde{\theta}\ \ \ \&\ \ \ G_{2}=\cos4\tilde{\theta}+2
\label{eq:local-invariants}
\end{eqnarray}
and as a result the corresponding points on a 3-torus geometric structure \cite{balakrishnan09} 
\begin{eqnarray}
(c_{1}, c_{2}, c_{3})=(2\tilde{\theta}, 0, 0). 
\label{eq:WC-coordinate}
\end{eqnarray}
Consequently, we obtained the entangling power \cite{zanardi00-r, balakrishnan09} 
\begin{eqnarray}
e_{P}[U(\tilde{\mathcal{C}})]=\frac{2}{9}\sin^{2}2\tilde{\theta}.
\label{eq:entangling-power}
\end{eqnarray}
Figure \ref{fig:EP-WC} shows the geometric 3-torus points $(c_{1}, c_{2}, c_{3})$ in the form of symmetry reduced coordinates on the tetrahedron Weyl chamber \cite{balakrishnan09} as well as the entangling power. 

It is important to note that, all the entangling characteristics of the $U(\tilde{\mathcal{C}})$ given by local invariants in Eq. (\ref{eq:local-invariants}), geometric 3-torus points in Eq. (\ref{eq:WC-coordinate}), and the entangling power in Eq. (\ref{eq:entangling-power}) depend only on the ratio between the exchange couplings $J_{32}$ and $J_{42}$ represented, in fact, by the angle $\tilde{\theta}=2\arctan\frac{J_{32}}{J_{42}}$. The above analysis and Fig. \ref{fig:EP-WC} indicate that the geometric two-qubit gate $U(\tilde{\mathcal{C}})$ is an entangling gate for any $0<\tilde{\theta}<\pi/2$. Moreover, for $\tilde{\theta}=\pi/4$, the $U(\tilde{\mathcal{C}})$ is a special perfect entangler \footnote{A special perfect entangler is a perfect entangler that can maximally entangle a full product basis \cite{rezakhani03}}, which belongs to the CNOT equivalence class represented by the point $L$ on the Weyl chamber and owns the maximum gate entangling power of $2/9$.

\begin{figure}[h]
\centering
\includegraphics[width=60mm,height=40mm]{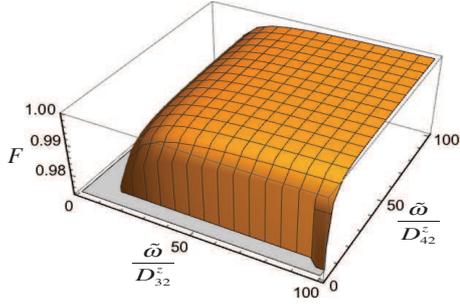}
\caption{(Color online) Fidelity, $F$, of the geometric two-qubit  entangling gate $U(\tilde{\mathcal{C}}; \tilde{\theta}=\pi/4)$ against dimensionless parameters $\frac{\tilde{\omega}}{D^{z}_{32}}$ and $\frac{\tilde{\omega}}{D^{z}_{42}}$ corresponding to the Dzyalozhinsky-Moriya spin-orbit interaction contribution to the system. We have used the antisymmetric Dzyalozhinsky-Moriya term $\frac{D^{z}_{32}}{2}[\sigma_{x}^{(3)}\sigma_{y}^{(2)}-\sigma_{y}^{(3)}\sigma_{x}^{(2)}]+\frac{D^{z}_{42}}{2}[\sigma_{x}^{(2)}\sigma_{y}^{(4)}-\sigma_{y}^{(2)}\sigma_{x}^{(4)}]$.}
\label{fig:Fid-2QEG}
\end{figure}

Similar to the case of single-qubit gate, we implement the geometric two-qubit entangling gate $U(\tilde{\mathcal{C}})$ through only an anisotropic XY interaction coupling and consider the antisymmetric Dxzialoshinsky-Moriya contribution as a perturbative noise to this XY coupling. Figure \ref{fig:Fid-2QEG} shows the fidelity of the geometric two-qubit  
spacial perfect entangler $U(\tilde{\mathcal{C}}; \tilde{\theta}=\pi/4)$ against this noise. 

\section{summary}
\label{summary}
In summary, we have introduced a practical procedure to implement universal nonadiabatic holonomic quantum information processing through decoherence-free subspaces. We have used two physical qubits to encode a logical qubit into a two dimensional decoherence-free subspace. A universal set of nonadiabatic holonomic single-qubit gates is then obtained by embedding the logical qubit into a three dimensional decoherence-free subspace and constructing a lambda structure. This is done by coupling the two physical qubits through an auxiliary third qubit in an anisotropic XY interaction Hamiltonian. In order to perform universal quantum computation, we have also demonstrated a family of entangling nonadiabatic holonomic operations on two logical qubits, which includes entangling gate equivalent to CNOT gate. The entangling gates are achieved by forming a double lambda structure in an anisotropic XY interaction coupling as well. The proposed scheme differs from previous proposals in that the entire procedure here is based on typical one dimensional anisotropic XY Hamiltonian, which can be fabricated in different physical systems such as NMR \cite{niederberger2010, alvarez2010, rao2014}, quantum dot spins \cite{imamoglu1999}, atomic systems \cite{barredo2015}, nitrogen-vacancy centers in diamond \cite{ping2013}, magnetic impurities on the surface of topological insulators \cite{zhu2011}, and optical systems \cite{struck2011, kosior2013, berloff2017, tamate2016, takeda2017}. This in turn adds to the feasibility of the present proposal.

\section*{Acknowledgments}
The author acknowledges support from Department of Mathematics at University of Isfahan
(Iran).

\bibliography{NHQCDF}

\end{document}